\documentclass[12pt]{article}
\input epsf.sty
\usepackage{mathrsfs}
\usepackage{amsmath}
\usepackage{mathrsfs}
\usepackage{amssymb}
\usepackage{color}
\usepackage{psfrag}
\usepackage{graphicx}
\usepackage{subfigure}
\usepackage{rotating}
\usepackage{cite}

\newcommand{\bea}{\begin{eqnarray}}
\newcommand{\eea}{\end{eqnarray}}
\newcommand{\be}{\begin{equation}}
\newcommand{\ee}{\end{equation}}
\newcommand{\vs}[1]{\vspace{#1 mm}}

\newcommand{\dsl}{\pa \kern-0.5em /}

\newcommand{\pa}{\partial}

\newcommand{\nn}{\nonumber\\}

\begin{document}
\topmargin 0pt
\oddsidemargin 0mm

\begin{flushright}



\end{flushright}

\vspace{2mm}

\begin{center}
{\Large \bf Entanglement thermodynamics for an excited state of Lifshitz system}

\vs{10}

{Somdeb Chakraborty\footnote{E-mail: somdeb.chakraborty@saha.ac.in}, Parijat Dey\footnote{E-mail: parijat.dey@saha.ac.in},
Sourav Karar\footnote{E-mail: sourav.karar@saha.ac.in} and  
Shibaji Roy\footnote{E-mail: shibaji.roy@saha.ac.in}}

 \vspace{4mm}

{\em

 Saha Institute of Nuclear Physics,
 1/AF Bidhannagar, Calcutta-700 064, India}

\end{center}

\vs{10}

\begin{abstract}

\end{abstract}
A class of (2+1)-dimensional quantum many body system characterized by an anisotropic scaling
symmetry (Lifshitz symmetry) near their quantum critical point can be described by a (3+1)-dimensional dual gravity 
theory with negative cosmological constant along with a massive vector field, where the scaling symmetry is realized
by the metric as an isometry. We calculate the entanglement entropy of an excited state of such a system holographically,
i.e., from the asymptotic perturbation of the gravity dual using the prescription of Ryu and Takayanagi, when the
subsystem is sufficiently small. With suitable identifications, we show that this entanglement entropy satisfies
an energy conservation relation analogous to the first law of thermodynamics. The non-trivial massive vector field here
plays a crucial role and contributes to an additional term in the energy relation.
  
\newpage

\noindent{\it 1. Introduction} : When a system is in thermal equilibrium, it can very well be described by a few
macroscopic quantities, like, temperature, pressure, volume, energy, entropy and various chemical potentials associated with
conserved charges of the system. The system is governed by certain laws called the thermodynamical laws by which these 
quantities are related to each other. One such law is the first law of thermodynamics which simply describes the conservation
of energy of the system, in particular, it relates the change in energy $(dE$) to the change in entropy ($dS$) or the amount 
of information stored in the system by a proportionality constant, the temperature, i.e., $dE=TdS$. 

It is, however, not so easy to describe a system when it is away from equilibrium. For a quantum system even if it is away
from equilibrium, the quantum information can be encoded in a quantity called entanglement entropy  (see \cite{Bombelli:1986rw, 
Srednicki:1993im,
Holzhey:1994we, Calabrese:2004eu, Calabrese:2005zw, Calabrese:2009qy, Eisert:2008ur}). The energy of the system
can also be given independently whether it is at or away from equilibrium. Then it is interesting to ask whether an analogous
thermodynamical first law like relation holds good even when the system is away from equilibrium. This question was addressed 
\cite{Bhattacharya:2012mi}
for some excited state (also see \cite{Wong:2013gua,Ben-Ami:2014gsa} for related works which compute the holographic entanglement entropy for excited states) of certain quantum system and then studying the relation of the change in entanglement entropy obtained 
by holographic method \cite{Ryu:2006bv,Ryu:2006ef} and the corresponding change in energy of 
the excited state of the system. It was indeed found that for 
sufficiently small subsystem, the change in entanglement entropy is proportional to the change in energy, where the proportionality
constant is related to the size of the entangling region. Identifying the proportionality constant as the inverse of some
temperature, the so-called entanglement temperature, we get a first law in analogy with first law of thermodynamics.

The simple first law where the change in entanglement entropy is directly proportional to the change in energy is not necessarily
true always. It is true for the case studied in \cite{Bhattacharya:2012mi}, where it was the consequence of the choice of 
translationally and rotationally
invariant excited state. However, for rotationally non-invariant (anisotropic) excited states the change in entanglement entropy
will not only involve change in energy, but it will also involve a pressure-like term as obtained in 
\cite{Guo:2013aca,Allahbakhshi:2013rda}. So, defining
a new pressure-like quantity called the entanglement pressure one can still get a modified first law in analogy with thermodynamical
first law.

However, in all the cases the first law of entanglement thermodynamics were obtained by studying the relativistic system. 
In this paper we look at the excited state of a non-relativistic system, the Lifshitz system in four dimensions,
where there is no conformal invariance. Not only that, the excited state we consider will be non-isotropic. So, the change in entropy
will involve apart from an entanglement pressure term an additional term which is related to the entangling region and in analogy
with the first law of thermodynamics can be identified with an entanglement chemical potential and a first law-like relation can be written.
To be precise, we start from a dual gravity solution in four space-time dimensions which admits a Lifshitz scaling symmetry. The 
solution follows from a four dimensional gravity action with a cosmological constant and a massive gauge field. We consider an
excited state corresponding to an asymptotic Lifshitz solution obtained in \cite{Ross:2009ar} by perturbation and solving the linearized equations
of motion in the radial gauge. We compute the entanglement entropy of this excited state on a strip-type subsystem using holographic method
and find that it is finite and contains three terms proportional to energy, pressure and a term which can be identified with a chemical 
potential. With proper identifications we recover a first law-like relation - a first law of entanglement thermodynamics. In the relativistic
limit we find that the chemical potential term vanishes and we recover the relativistic result obtained before when the space-time dimension
is four.

This paper is organized as follows. In the next section we give the relevant part of the holographic stress tensor complex for
a four dimensional asymptotically Lifshitz theory as discussed in detail in \cite{Ross:2009ar}. In section 3, we compute the holographic
entanglement entropy of this excited state of the Lifshitz system and obtain the first law of entanglement thermodynamics.
Then we conclude in section 4.
  
\vspace{.5cm}

\noindent{\it 2. Lifshitz theory and holographic stress tensor} : In this section we will briefly review the asymptotic
perturbation of four dimensional Lifshitz theory and the associated holographic stress tensor of the theory \cite{Ross:2009ar}. 
We here mention the relevant part of the stress tensor complex (required for obtaining the first law of entanglement 
thermodynamics in the next section) for constant perturbation of four dimensional Lifshitz theory.
It is well-known that a four dimensional Lifshitz metric can be obtained from a bulk action (see \cite{Kachru:2008yh, Taylor:2008tg} ) of the form,
\be\label{action}
S = \frac{1}{16\pi G_4}\int d^4x \sqrt{-g}\left(R - 2\Lambda - \frac{1}{4}F_{\mu\nu}F^{\mu\nu}
- \frac{1}{2} m^2 A_\mu A^\mu\right)
\ee
where $A_\mu$ is a massive gauge field and $\Lambda$ is a cosmological constant. The equations of motion 
following from the above action are,
\bea\label{eom}
R_{\mu\nu} &=& \Lambda g_{\mu\nu} + \frac{1}{2} F_{\mu\rho} F_{\nu}^{\,\,\rho} - \frac{1}{8} F_{\rho\sigma} F^{\rho\sigma}g_{\mu\nu}
+ \frac{1}{2} m^2 A_\mu A_\nu\nn
\nabla_\mu F^{\mu\nu} &=& m^2 A^\nu
\eea  
Choosing $\Lambda = -\frac{1}{2}(z^2+z+4)$ and $m^2 = 2z$, the equations of motion \eqref{eom} have a solution of 
the form \cite{Ross:2009ar}
\bea \label{lmet}
ds^2 &=& - r^{2z} dt^2 + r^2 (dx^2 + dy^2) + \frac{dr^2}{r^2}\nn
 \quad A &=& \alpha r^z dt ,\quad \alpha^2 = \frac{2(z-1)}{z}.
\eea
The solution has a Lifshitz scaling symmetry : $t \to
\lambda^z t,\, x \to \lambda x,\, y \to \lambda y,\,  r \to \lambda^{-1} r$, where $z$ is the dynamical scaling exponent and $\alpha$
is a constant as given above. The solution \eqref{lmet} is the gravity dual of a (2+1) dimensional quantum many body system with a 
Lifshitz symmetry near its quantum critical point. A complete action including a boundary term which is on-shell finite and leads to
a well-defined action principle is given as,
\bea\label{actiontotal}
S_{\rm tot} &=&  \frac{1}{16\pi G_4}\int d^4x \sqrt{-g}\left(R - 2\Lambda - \frac{1}{4}F_{\mu\nu}F^{\mu\nu}
- \frac{1}{2} m^2 A_\mu A^\mu\right)\nn
&& \qquad\qquad + \frac{1}{16\pi G_4}\int d^3\xi \sqrt{-h}\left(2K - 4 - z\alpha \sqrt{-A_\alpha A^\alpha}\right) + S_{\rm deriv}
\eea
where $\xi^\alpha$ are the boundary coordinates and $h_{\alpha\beta}$ is the induced metric on the boundary. $K = h^{\alpha\beta} K_{\alpha\beta}$ is the 
extrinsic curvature, where $K_{\alpha\beta}=\nabla_{(\alpha} n_{\beta)}$ with $n^\alpha$, a unit vector orthogonal to the boundary directed outwardly.
$S_{\rm deriv}$ are terms involving the derivatives of the boundary fields. We will ignore this term as it will not contribute for the constant
boundary fields in our case. The above action can then be used to define the stress tensor of the boundary theory by varying it with respect
to the boundary metric. Since here the boundary theory is non-relativistic, instead of a fully covariant stress tensor, one can define stress
tensor complex consisting of energy density, energy flux, momentum density and pressure density which satisfy the conservation relations.
For the calculation of entanglement entropy we will need only the energy and the pressure densities as we will see in the next section.
  
Now instead of the Lifshitz solution \eqref{lmet} which corresponds to the ground state we will consider an excited state by 
perturbing the solution as follows \cite{Ross:2009ar}, 
\bea \label{lmetpert}
  ds^2 &=& - r^{2z} (1+h_{tt}(r))dt^2 + r^2 (1+h_{xx}(r))dx^2 + r^2(1+h_{yy}(r))dy^2 + \frac{dr^2}{r^2}  \nn 
& & + 2 [-r^{2z} v_{1x}(r) + r^2 v_{2x}(r)] \; dt dx +2 [-r^{2z} v_{1y}(r) + r^2 v_{2y}(r)] \; dt dy + 2 r^2 h_{xy}(r)\; dx dy\nn
A &=& \alpha r^z [(1+ a_t(r) + \frac{1}{2}h_{tt}(r))dt + v_{1x}(r) dx + v_{1y}(r) dy] 
  \eea
Note that \eqref{lmetpert} gives an asymptotically Lifshitz space-time when 
$h_{tt}(r)$, $h_{xx}(r)$, $h_{yy}(r)$, $v_{1x}(r)$, $v_{2x}(r)$, $v_{1y}(r)$, $v_{2y}(r)$, $h_{xy}(r)$, and $a_t(r)$ go to zero as $r\to \infty$.
Here $h_{r\mu}=0$ gauge is chosen. Also we have assumed the perturbations to be constant in the boundary directions and so, $a_r=0$ by equation of
motion. Note that the perturbations have been split into scalar, vector and tensor parts and because of the rotational invariance of the
background in the $x$-$y$ plane they don't mix with each other.
Now defining
  \bea\label{defn}
  h_{tt}(r) &=& f(r)\nn
  h_{xx}(r) &=& k(r) + {t_d}(r) \nn
  h_{yy} (r) &=& k(r) - {t_d}(r) \nn
  a_t &=& j(r)
  \eea
the linearized action can be obtained by substituting \eqref{defn} in \eqref{actiontotal}. This action has been shown in \cite{Ross:2009ar} 
to be finite
on-shell and the various components of the stress tensor complex can be found by varying the action with respect to the boundary metric. 
We here give the expressions of the energy and pressure densities that we will need later. In terms of the functions defined above they 
have the forms \cite{Ross:2009ar},
\bea\label{energypressure}
T_{tt} &=& -r^{z+2}\left[2r\partial_r k(r) + \alpha^2\left(zj(r) + r\partial_rj(r) + \frac{1}{2} r \partial_r f(r)\right)\right]\nn
T_{xx} &=& -2r^{z+2}\left[(z-1)j(r) - \frac{r}{2}\partial_r f(r) - \frac{r}{2} \partial_r k(r) - \frac{1}{2} (z+2) t_d(r)\right]\nn
T_{yy} &=& -2r^{z+2}\left[(z-1)j(r) - \frac{r}{2}\partial_r f(r) - \frac{r}{2} \partial_r k(r) + \frac{1}{2}(z+2) t_d(r)\right]
\eea
We also need another function obtained from the linearized action by varying with respect to the massive gauge field and having
the form,
\be\label{szero}
s_0(r) = \alpha r^{z+2}\left[zj(r)+r\partial_r(\frac{1}{2}f(r) + j(r))\right]
\ee
This is not conserved but will be useful to express the extra term that appears in the entanglement entropy to be discussed in the next section.
The equations of motion following from the linearized action can be explicitly solved in the radial gauge and the   
normalizable solutions for $j(r), k(r), f(r)$ and $t_d(r)$ can be obtained. The solutions have quite different forms for $z=2$ and $z \neq 2$ and
therefore we give the solutions separately in the following \cite{Ross:2009ar, Ross:2011gu, Andrade:2013wsa}. For $z=2$,
\bea\label{z2}
 j(r) &=& -\frac{c_1 + c_2 \ln r}{r^4}, \nn
 f(r) &=& \frac{4 c_1 - 5 c_2 + 4 c_2 \ln r}{12 r^4}, \nn
 k(r) &=& \frac{4 c_1 + 5 c_2 + 4 c_2 \ln r}{24 r^4}, \nn
t_d(r) &=& \frac{t_{d2}}{r^4}
\eea
and for $z\neq 2$,
\bea\label{zneq2}
 j(r) &=& -\frac{(z+1)c_1}{(z-1)r^{z+2}} -
 \frac{(z+1)c_2}{(z-1)r^{\frac{1}{2}(z+2+\beta_z)}}, \nn
 f(r) &=& 4\frac{1}{(z+2)} \frac{c_1}{r^{z+2}} + 2
 \frac{(5z-2-\beta_z)}{(z+2+\beta_z)}
 \frac{c_2}{r^{\frac{1}{2}(z+2+\beta_z)}}, \nn
 k(r) &=& 2 \frac{1}{(z+2)} \frac{c_1}{r^{z+2}} - 2
 \frac{(3z-4-\beta_z)}{(z+2+\beta_z)}
 \frac{c_2}{r^{\frac{1}{2}(z+2+\beta_z)}},\nn
t_d(r) &=& \frac{t_{d2}}{r^{z+2}}  
 \eea
where $\beta_z^2 = 9z^2-20z+20 = (z+2)^2 +8(z-1)(z-2)$.
Note that in the above $c_1$, $c_2$ and $t_{d2}$ are integration constants. There are other integration constants in the solution which have
been put to zero, by coordinate redefinition and to recover asymptotic Lifshitz solution \eqref{lmet}. One of the constants
in \eqref{zneq2} involving term with a fall-off $r^{-\frac{1}{2}(z+2-\beta_z)}$ has been set to zero since it gives a divergent contribution for $z>2$ 
and it gives non-normalizable mode even for $z<2$ \cite{Andrade:2013wsa}. These solutions
can be used in \eqref{energypressure} to find the finite energy and the pressure densities (or their expectation values) in terms of the 
parameters or the constants of the
solution. We will also compute $s_0(r)$ using these solutions. Note that we have not given the solution for $v_{1x}(r)$, $v_{1y}(r)$, $v_{2x}(r)$
$v_{2y}(r)$ and $h_{xy}(r)$ explicitly since they will not be needed for the computation of the entanglement entropy of the excited state
which we turn to in the next section.

\vspace{.5cm}

\noindent{\it 3. Entanglement thermodynamics and a first law for the excited state} : In this section we will    
compute the (shift in the) holographic entanglement entropy of the excited state due to the metric perturbation 
\eqref{lmetpert} as given in \cite{Bhattacharya:2012mi, Bhattacharya:2013bna}. The entangling region is taken 
to be  a strip with width $\ell$ given by  
\be\label{strip}
 -\frac{\ell}{2}\leq
x\leq \frac{\ell}{2},\qquad\qquad 0\leq y \leq L.
\ee
The standard procedure \cite{Ryu:2006bv,Ryu:2006ef} of computing the holographic entanglement entropy is to minimize a 
surface embedded in the time
slice of the bulk geometry and ending at $r=\infty$ with the boundary coinciding with the entangling region. Then the entanglement
entropy would be given by the area of this minimal surface divided by $4G_4$. Parameterizing the surface by $x = x(r)$, we can calculate
the area from the induced metric (see \eqref{lmetpert}) as,
\be\label{area}
A = \int_{\infty}^{r_t} dr \int_0^L dy \sqrt{1+r^4x^{\prime\,2}(r)+r^4x^{\prime\,2}(r)(h_{xx}(r)+h_{yy}(r))+h_{yy}(r)}
\ee
Here `prime' denotes derivative with respect to $r$. Now minimizing the area determines the function $x(r)$ and substituting that
back into \eqref{area} we finally obtain,
\be\label{area1} 
A = A^{(0)} + \frac{L}{2}\int_{r_t}^\infty dr \frac{h_{yy}(r) + \left(\frac{r_t}{r}\right)^4 h_{xx}(r)}
{\sqrt{1- \left(\frac{r_t}{r}\right)^4}}
\ee
and
\be\label{ell}
\ell =  \frac{2}{{r}_t}\int_1^{\infty} \frac{d\zeta}{\zeta^2\sqrt{\zeta^4-1}} =  \frac{2 \sqrt{\pi} \; \Gamma\left(\frac{3}{4}\right)}
{\Gamma\left(\frac{1}{4}\right) \; r_t}
\ee
where $r_t$ is the turning point and $\zeta = r/r_t$. Also here $A^{(0)}$ denotes the area of the minimal surface for the Lifshitz background.
Since we are considering small perturbations around this background, $h$'s are small such that the expansion in \eqref{area1} makes
sense. This in turn implies that the turning point $r_t$ is actually close to the boundary and we do not bother about the IR geometry.
Now since from \eqref{ell} $r_t$ is inversely related to the measure of the entangling region, i.e., $\ell$, our analysis is valid
for very small subsystem. From \eqref{area1} we get the shift in area $\Delta A$ due to the perturbation
which in turn will give us the shift in entanglement entropy of the excited state over the pure Lifshitz ground state as,
\be\label{dels}
\Delta S = \frac{L}{8G_4} \int_{r_t}^{\infty} dr \; \frac{1}{\sqrt{1-{(\frac{r_t}{r})}^4}} \Big[h_{yy}(r)+ {\Big(\frac{r_t}{r}\Big)}^4 h_{xx}(r)\Big]
\ee
In \eqref{defn} $h_{xx}(r)$ and $h_{yy}(r)$ are defined in terms of the functions $k(r)$ and $t_d(r)$, whose solutions are given in \eqref{z2}
for $z=2$ and in \eqref{zneq2} for $z\neq 2$. We will therefore evaluate $\Delta S$ in the following for the two cases separately.
 
For $z=2$, using the form of $k(r)$ and $t_d(r)$ as given in \eqref{z2}, \eqref{dels} reduces to
\be\label{delsf}
 \Delta S = \frac{L \sqrt{\pi}  \Gamma(\frac{3}{4})}{8 G_4\Gamma(\frac{1}{4}) 24 {r_t}^3}\Big[\frac{32 c_1}{5}- 
\frac{48 t_{d2}}{5} + c_2 (\frac{352}{25}-\frac{8 \pi}{5}) + \frac{32}{5} c_2 \ln r_t\Big]
\ee
Now using \eqref{ell} we will replace one $r_t$ in the denominator of the $\Delta S$ expression in \eqref{delsf} and write it as, 
\be\label{delszeq2}
\Delta S = \frac{\ell L \pi}{24 {r_t}^2} \frac{1}{16 \pi G_4} \Big[ \frac{32 c_1}{5} - 
\frac{48 t_{d2}}{5} + c_2 (\frac{352}{25} -\frac{8 \pi}{5}) + \frac{32}{5}c_2\ln r_t\Big]
\ee
To express $\Delta S$ in terms of the holographic energy-momentum tensors, we first write down their forms \eqref{energypressure}
using various functions given in \eqref{z2} and for $z=2$ we have \cite{Ross:2009ar}, 
\bea\label{emtensor1}
\langle T_{tt} \rangle  &=& \frac{1}{16 \pi G_4} {\frac{4 c_2}{3}}\nn
\langle T_{xx} \rangle &=& \frac{1}{16 \pi G_4} \Big({\frac{4 c_2}{3}+4 t_{d2}} \Big).
\eea
Using these in \eqref{delszeq2} we get,
\be\label{deltas}
\Delta S = \frac{\pi}{24 r_t^2}\frac{(324-30\pi)}{25}\left[\langle T_{tt} \rangle L\ell - \frac{60}{(324-30\pi)}
\langle T_{xx} \rangle L\ell + \frac{160}{(324-30\pi)} 
\frac{L\ell}{16\pi G_4}(c_1 + c_2 \ln r_t)\right]
\ee
Defining an entanglement temperature as \cite{Bhattacharya:2012mi,Bhattacharya:2013bna,Allahbakhshi:2013rda},
\be
T_{\rm ent} = \frac{24 r_t^2}{\pi}\frac{25}{(324-30\pi)} = \frac{96 \Gamma^2\left(\frac{3}{4}\right)}{\ell^2 \Gamma^2\left(\frac{1}{4}\right)}
\frac{25}{(324-30\pi)}
\ee
where we have used \eqref{ell} in the second equality and also the total energy and entanglement pressure as,  
\bea
  \Delta E &=& \int_{0}^{L} \int_{-\ell/2}^{ \ell/2} dy\; dx\;   \langle T_{tt} \rangle  \nn &=& L\; \ell \;\langle T_{tt} \rangle \nn
  \Delta P_x &=&  \langle T_{xx} \rangle 
  \eea
we can rewrite \eqref{deltas} as,
\be\label{thermo}
\Delta E = T_{\rm ent} \Delta S + \frac{60}{(324-30\pi)}\Delta P_x V - \frac{160}{(324-30\pi)}\frac{L\ell}{16\pi G_N}(c_1 + c_2 \ln r_t),
\ee
where $V=L\ell$ is the volume of the entangling region.
This looks like the first law of entanglement thermodynamics modulo the last term. We note that the stress tensor complex for the 
Lifshitz theory has contribution not only from the metric but also from the gauge field. The gauge field part is actually encoded 
in the function $s_0(r)$ given in \eqref{szero}. This quantity by itself is not conserved and is dual to the gauge field $A^0 = r^{-z}A_t 
= \alpha$. We can 
evaluate $\langle s_0(r) \rangle$ for $z=2$ with the solution of the functions given in eq.\eqref{z2} and we obtain,
\be 
\langle s_0(r) \rangle = \frac{4}{3}(c_1 + c_2 \ln r)
\ee
precisely the combination we have in the last term of our expression of $\Delta E$ in \eqref{thermo}. Also note that the quantity
$\langle s_0 \rangle/(16\pi G_4)$ has the same scaling dimension as the energy density or the chemical potential. Also if we evaluate 
$\langle s_0(r) \rangle$ at the 
turning point $r_t$, then it can be interpreted as a physical quantity at the boundary by the relation \eqref{ell}. Thus identifying 
\be
\Delta \mu \equiv \frac{\langle s_0(r_t)\rangle}{16 \pi G_4}
\ee
where $\Delta \mu$ is the entanglement chemical potential of the excited state, we can write the first law of the entanglement 
thermodynamics as,
\be\label{thermofinal}
\Delta E = T_{\rm ent} \Delta S + \frac{60}{(324-30\pi)}\Delta P_x V - \frac{30}{(324-30\pi)}\Delta \mu Q.
\ee
Note that we have used $Q=m^2\alpha V$ in the last term, with $m$ and $\alpha$ as defined in eqs. \eqref{lmet} and \eqref{eom}.
Also we have identified $m^2A^0=m^2\alpha$ as some charge density $j^0$ using the equation of motion \eqref{eom}. Thus, 
\eqref{thermofinal} represents a modified first law of entanglement thermodynamics for the Lifshitz system for $z=2$.

The shift in holographic entanglement entropy can similarly be calculated for $ z \neq 2$ from \eqref{dels} using the forms
of $h_{xx}(r) = k(r) + t_d(r)$ and $h_{yy}(r) = k(r) - t_d(r)$ as given in \eqref{zneq2} and so we have,
\be \label{delsfinal}
\Delta S  = \frac{ L \sqrt{\pi} }{16   G_4 {r_t}^{z+1}} \Big[ \frac{\Gamma(\frac{1+z}{4})}{\Gamma(\frac{3+z}{4})} 
\frac{1}{(z+3)} (2 c_1 -t_{d2}) +    {r_t}^{{\frac{1}{2}}{(z+2- \beta_z)}} \frac{\Gamma(\frac{z+ \beta_z}{8})}
{\Gamma(\frac{{z+4+ \beta_z}}{8})} \frac{2 (4+\beta_z -3z)}{4+z+ \beta_z} c_2\Big]
\ee
Again we replace one $r_t$ in the denominator of \eqref{delsfinal} in terms of $ \ell $ and get,
 \be\label{delszneq2}
 \Delta S = \frac{L  \ell \Gamma(\frac{1}{4}) \pi}{2\Gamma(\frac{3}{4}) {r_t}^z}\frac{1}{16 \pi G_4} \Big[ \frac{\Gamma(\frac{1+z}{4})}
{\Gamma(\frac{3+z}{4})} \frac{1}{(z+3)} (2 c_1 -t_{d2})  +  {r_t}^{{\frac{1}{2}}{(z+2- \beta_z)}} \frac{\Gamma(\frac{z+ \beta_z}{8})}
{\Gamma(\frac{{z+4+ \beta_z}}{8})} \frac{2 (4+\beta_z -3z)}{4+z+ \beta_z} c_2\Big].
\ee
We will express $\Delta S$ in terms of the energy-momentum tensor and the chemical potential as before.
The holographic stress energy tensor for $z \neq 2$ can be calculated as before from \eqref{energypressure} using the functions given
in \eqref{zneq2} and they are given as \cite{Ross:2009ar},
\bea\label{energypress}
\langle T_{tt} \rangle  &=&  \frac{1}{16 \pi G_4} {\frac{4 (z-2)}{z} c_1} \nn
\langle T_{xx} \rangle &=&  \frac{1}{16 \pi G_4}[{2(z-2)c_1 +(z+2) t_{d2}}].
\eea
The chemical potential in this case can be obtained from $s_0(r)$ given in \eqref{szero} with the functions given in \eqref{zneq2}
and has the form, 
\be\label{chemical2}
\Delta \mu = \frac{1}{16\pi G_4} \langle s_0(r_t) \rangle = \frac{1}{16\pi G_4}\frac{\alpha}{z-1}\left[4c_1 + c_2 z(4+\beta_z - 3z) 
r_t^{\frac{1}{2}(z+2-\beta_z)}\right]
\ee
The relations \eqref{energypress} and \eqref{chemical2} can be inverted to obtain,
\bea\label{inverted}
c_1 &=& \frac{16\pi G_4 z}{4(z-2)} \langle T_{tt} \rangle,\nn
t_{d2} &=& \frac{16\pi G_4(2\langle T_{xx} \rangle - z \langle T_{tt}) \rangle}{2(z+2)},\nn
c_2 &=& \frac{16\pi G_4\left[(z-2)(z-1) \Delta \mu - \alpha z \langle T_{tt} \rangle\right]}{\alpha z(z-2)(4+\beta_z-3z)r_t^{\frac{1}{2}(z+2-\beta_z)}}
\eea
Now using \eqref{inverted} in \eqref{delszneq2} we get,
\be\label{thermo2}
\Delta S = \frac{\pi \Gamma\left(\frac{1}{4}\right) A_1}{2 r_t^z \Gamma\left(\frac{3}{4}\right)}\left[\langle T_{tt} \rangle L\ell - \frac{A_2}{A_1} 
\langle T_{xx} \rangle L\ell +
\frac{A_3}{A_1} \Delta \mu (m^2\alpha L\ell)\right]
\ee
where
\bea
A_1 &=& \frac{z^2}{(z+3)(z^2-4)}\frac{\Gamma\left(\frac{1+z}{4}\right)}{\Gamma\left(\frac{3+z}{4}\right)} - \frac{2}{(z-2)(4+z+\beta_z)}
\frac{\Gamma\left(\frac{z+\beta_z}{8}\right)}{\Gamma\left(\frac{z+4+\beta_z}{8}\right)},\nn
A_2 &=& \frac{1}{(z+3)(z+2)}\frac{\Gamma\left(\frac{1+z}{4}\right)}{\Gamma\left(\frac{3+z}{4}\right)}, \quad
A_3 \,\,=\,\, \frac{1}{2 z (4+z+\beta_z)}\frac{\Gamma\left(\frac{z+\beta_z}{8}\right)}{\Gamma\left(\frac{z+4+\beta_z}{8}\right)}
\eea 
Now defining the entanglement temperature as \cite{Bhattacharya:2012mi,Bhattacharya:2013bna,Allahbakhshi:2013rda},
\be\label{temp}
T_{\rm ent} = \frac{2 r_t^z \Gamma\left(\frac{3}{4}\right)}{\pi \Gamma\left(\frac{1}{4}\right) A_1}
\ee
we can rewrite \eqref{thermo2} as,
\be\label{thermo3}
\Delta E = T_{\rm ent} \Delta S + \frac{A_2}{A_1} \Delta P_x V - \frac{A_3}{A_1} \Delta \mu Q
\ee
where we have used $Q=m^2\alpha V$.
This is the modified first law of entanglement thermodynamics \cite{Allahbakhshi:2013rda} for Lifshitz system with $z\neq 2$. Note that
in the expression of $\Delta E$ (both for $z=2$ and $z\neq 2$) only the $x$-component of the pressure, i.e., the component normal to the 
entangling region appears. If 
we try to include the total pressure $P_x + P_y$, some of the parameters in the $\Delta S$ expression in \eqref{deltas} and \eqref{delszneq2} 
can not be eliminated and we don't get a first law like relation. 

Let us see how we can recover the AdS or relativistic result from here. The relativistic case can be obtained when $z=1$. We note that
for $z=1$, $Q=0$ and therefore, the last term in \eqref{thermo3} vanishes as expected. Also, note that
for $z=1$, $A_1=\sqrt{\pi}/6$ and $A_2 = \sqrt{\pi}/12$ and therefore, the coefficient of $\Delta P_x V$ term is $A_2/A_1 = 1/2$ which can be compared with eq.(2.33) of 
\cite{Allahbakhshi:2013rda}, where the coefficient has the form $(d-1)/(d+1)$ which is precisely 1/2 for $d=3$ as in our case. Finally we
can check the form of entanglement temperature given in \eqref{temp} with that given in \cite{Allahbakhshi:2013rda}. For $z=1$, we
get from \eqref{temp} $T_{\rm ent} = 24 \Gamma^2(3/4)/(\pi \ell \Gamma^2(1/4))$ which differs by a factor of 2 with that given in 
ref.\cite{Allahbakhshi:2013rda}.
   
\vspace{.5cm}

\noindent{\it 4. Conclusion} : To conclude, in this paper we have holographically computed the change in entanglement entropy
for the excitation of the ground state Lifshitz system. The gravity dual of the excited state is given by the asymptotic perturbation
of the (3+1) dimensional Lifshitz solution. The holographic entanglement entropy of the excited state is calculated for a very small
strip-type subsystem from the metric by the standard method of Ryu and Takayanagi. We have used the stress energy complex of the above mentioned 
asymptotic Lifshitz solution obtained in \cite{Ross:2009ar} and tried to express the change in entanglement entropy in terms of change 
in energy and change in entanglement pressure. We found that unlike the relativistic system with conformal symmetry, the change in 
entanglement entropy for the Lifshitz system contains an additional term. We have attributed the origin of the additional term to the
presence of the massive gauge field and identified it, in analogy with thermodynamical first law, with the change in entanglement chemical 
potential, which depends on the entangling region, times a charge. We thus obtain a non-relativistic modification of first law of entanglement 
thermodynamics. When conformal symmetry is restored or in the relativistic case, i.e., when the dynamical exponent $z \to 1$, our result
reduces to the first law obtained before for the AdS case in four dimensions. A similar calculation for the entanglement entropy can be
done for the spherical entangling region to obtain the non-relativistic first law of entanglement thermodynamics. However, it will not
give any new physical content, only the coefficients of the various terms will change due to the change in geometry of the entangling region.
One can also consider the general perturbations instead of the constant perturbations of the background and see how this changes the first law
of entanglement thermodynamics. Finally, it would also be of interest to extend our analysis to more general backgrounds like Lifshitz black 
hole and background having Lifshitz symmetry with hyperscaling violation.

\vspace{.5cm}

\noindent{\it Acknowledgements} : We would like to thank Simon F Ross for numerous communications and explanations on various aspects
of \cite{Ross:2009ar}. We are also grateful to Tadashi Takayanagi and Mohsen Alishahiha for useful comments on this work. One of us    
(PD) would like to acknowledge thankfully the financial support of the Council of Scientific and Industrial Research, India
(SPM-07/489 (0089)/2010-EMR-I).

\end{document}